\documentclass[twocolumn,showpacs,amsmath,amssymb]{revtex4}

\usepackage{graphicx}
\usepackage{dcolumn}
\usepackage{bm}

\def\ket#1{\mid #1\,\rangle}
\def\bra#1{\langle\,#1\mid}

\begin{document}

\title{On the entanglement of degenerated ground state for spin 1 and 1/2 pair}
\author{You-Quan Li and Guo-Qiang Zhu, Xue-An Zhao}
\affiliation{Zhejiang Institute of Modern Physics,
Zhejiang University, Hangzhou 310027, P.R. China}

\received{24 February 2004, revised 20 August 2004}

\begin{abstract}
We study the entanglement feature of the ground state of a system composed of
spin 1 and 1/2 parts. The concurrence vector is shown to
be consistent with the measurement of von Neumann entropy for such system.
In the light of the ground state degeneracy, we suggest a
\emph{average concurrence} to measure the entanglement of Hilbert
subspace. The entanglement property of both a general superposition as well as the
mixture of the degenerated ground states are discussed by means of
average concurrence and the negativity respectively.
\end{abstract}


\maketitle

\section{Introduction}

Entanglement is a fundamental concept in the theory of quantum
information, and the essential resource for modern applications of
quantum mechanics. In recent years, much attention have been paid
to the study of the entanglement of quantum systems either
qualitatively or  quantitatively. Entanglement possess some
resemblance to classical correlation, but it differs in some key
aspects, e.g., the entangled objects can violate Bell's
inequality\cite{bell,nl,dirk}. An useful measure of entanglement
for bipartite systems is the entanglement of formation. For a pure
state $\ket{\psi_{AB}}$, the entanglement of formation is given by
the entropy of the marginal density operator $\rho_{A}$ and
$\rho_{B}$ for system $A$ and $B$. In 1997, Wootters and Hill
introduced a new quantity called \emph{concurrence} that measures
the entanglement of system of qubits\cite{hill}
\begin{equation}
C_{\ket{\psi}}=|\bra{\psi}\sigma_{y}\otimes\sigma_{y}\ket{\psi^{*}}|
\nonumber
\end{equation}
Later, Wooters showed that the entanglement of
formation of an arbitrary two-qubit mixed state $\rho_{AB}$ can be
written in terms of the minimum average pure-state concurrence and
he derived an explicit expression for this minimum average
pure-state concurrence. He called this minimum the concurrence of
the mixed state\cite{wt}. After their work, many authors tried to
extend this strategy\cite{ru,pi}. For example, Rungta \emph{et al}
made an attempt to generalize the notion of concurrence to pure
bipartite states in arbitrary dimension by introducing operation
of universal inverter, which acts on quantum system of arbitrary
dimension\cite{ru}. They got one new extension of concurrence as
$C^{2}(\Psi)=2\nu_{D_{1}}\nu_{D_{2}}[1-tr(\rho_{A}^{2})]$. Let
$\nu_{D_{1}}=\nu_{D_{2}}=1$, the concurrence arranges from 0 to
$\sqrt{2(D-1)/D}$, where $D=min\{D_{1},D_{2}\}$. The so-called I
concurrence measures the entanglement of a pure state in terms of
the purity, $tr(\rho_{A}^{2})=tr(\rho_{B}^{2})$, of the marginal
density operators. Lozinski \emph{et al}\cite{loz} studied the
entanglement of $2\times K$ quantum system, they derived a
analytical expression for the low bound of the concurrence of
mixed quantum states of composite $2\times K$ system. Recently we
proposed a convenient generalization, \emph{concurrence vector} to
measure systems with higher symmetry.

It is known that the spin model is an important model in condensed
matter physics and statistical mechanics. Very recently are there
substantial discussions on the entanglement of quantum spin system
in equilibrium states. Particularly, both the entanglement of a
ground state and that of a thermal state of spin-$\frac{1}{2}$
pair with Heisenberg-type coupling have been investigated. Some
authors studied the thermal entanglement of qubit pairs in
the presence of a magnetic field  and found some interesting
results \cite{xgw,chenhong}. O'Connor and Wootters considered the
entangled chain with translational invariance \cite{wootters} and
studied how large one can make the nearest neighbor entanglement.
The pair entanglement and quantum phase transitions for XXZ model~\cite{GuLinLi}
and the local entanglement and quantum phase transition for extended Hubbard model
~\cite{GuLiLin} were studied in detail.

All the mentioned  investigations will undoubtedly set up a bridge connecting the quantum
information theory and condensed matter physics. However, the entanglement properties of
the ground state of a physical model has not been well defined. Particularly, the ground
state maybe degenerated, which request us to measure the entanglement of a
Hilbert subspace. In present paper, we study the entanglement feature of the
ground state for the systems composed of spin-$1/2$ and spin-$1$
parts. In Sec.~\ref{sec:reliable}, we show that the concurrence
vector is a reliable measurement of entanglement for spin-$1$ and spin-$1/2$
system. In Sec.~\ref{sec:average}, we propose a concept,
\emph{average concurrence} to measure the entanglement
of Hilbert subspace. On the basis of this concept, we discuss the entanglement
property of that system in Sec.~\ref{sec:superposition} for the general
superposition of the degenerate ground states. In Sec.~\ref{sec:mixture},
the general mixture of the degenerate ground states are discussed in terms of
Negativity \cite{vidal}. A brief summary with discussion is given in
the last section.

\section{Concurrence vector and its reliability}\label{sec:reliable}

We recently extended the concurrence originally proposed by Hill and
Wootters to a concurrence vector \cite{LiZhu}
\begin{equation*}
\textbf{C}=
   \{\bra{\psi}(E_{\alpha}-E_{-\alpha})\otimes(E_{\beta}-E_{-\beta})\ket{\psi^*}
    |\alpha,\!\beta\in\!\Delta^{+}\}
\end{equation*}
where $\Delta^{+}$ denotes the set of positive roots of $A_{N-1}$
Lie algebra. The above concurrence vector can be used to measure
the entanglement of high-dimensional systems.

For a pair of qubit and qutrit which can be
regarded as spin-$1/2$ and spin-1, the concurrence
vector is a three dimensional vector given by
\begin{equation}\label{a}
\textbf{C}=\{\bra{\psi}(\sigma_{+}-\sigma_{-})
  \otimes(E_{\alpha}-E_{-\alpha})\ket{\psi^*}
   \,\mid\alpha\in\Delta^{+}\}
\end{equation}
where
$\sigma_{\pm}=(\sigma_{x}\pm i\sigma_{y})/2$, and
$\Delta^{+}$ for $A_{2}$
contains three positive roots. Thus the concurrence vector here is of three dimension.

In order to show the reliability of concurrence vector, we consider
the von Neumann entropy of the system consisting of spin-1/2 and spin-1 parts.
As we known, any state of  bipartite system can be expanded as
\begin{equation}
|\psi\rangle=\sum_{\mu,j}a_{\mu j}|\mu\rangle\otimes|j\rangle,
\end{equation}
where $a_{\mu j}$ is complex coefficients, and in our present case,
$\mu=1, 2$ and $j=1, 2, 3$. The reduced density matrix
$\rho_{A}$ and $\rho_{B}$ can be easily obtained,
\begin{eqnarray*}
\rho_{A}=aa^{\dagger}=\left(\begin{array}{ccc}
  a_{11} & a_{12} & a_{13} \\
  a_{21} & a_{22} & a_{23} \\
\end{array}
\right)
\left(
\begin{array}{cc}
  a_{11}^{\ast} & a_{21}^{\ast} \\
  a_{12}^{\ast} & a_{22}^{\ast} \\
  a_{13}^{\ast} & a_{23}^{\ast} \\
\end{array}
\right).
\end{eqnarray*}
It is a $2\times2$ matrix, thus there are two eigenvalues
$\kappa_{1}^{2}$ and $\kappa_{2}^{2}$, that are squares of the coefficients
of Schmidt decomposition
$|\psi\rangle=\kappa_{1}|x_{1}\rangle_A |y_{1}\rangle_B
 +\kappa_{2}|x_{2}\rangle_A |y_{2}\rangle_B$.
Here the
$\kappa_{1}^{2}$ and
$\kappa_{2}^{2}$ are the
roots of the following secular equation
\begin{equation}
\lambda^2-\lambda+|\mathbf{C}|^{2}/4=0,
\label{aa}
\end{equation}
where $|\mathbf{C}|$ is precisely the norm of
concurrence vector we proposed, namely,
$
|\mathbf{C}|^2=C^2_{1}+C^2_{2}+C^2_{3}=
4(a_{11}a_{22}-a_{12}a_{21})^2
 +4(a_{12}a_{23}-a_{13}a_{22})^2+4(a_{11}a_{23}-a_{13}a_{21})^2
$.
From Eq.(\ref{aa}), we obtain
\begin{equation}\label{solution}
\kappa_{1,2}^2=\frac{1\pm\sqrt{1-|\mathbf{C}|^2}}{2}.
\end{equation}
So the von Neumann entropy is given by
\begin{equation}
E_N(|\psi\rangle)=h((1-\sqrt{1-|\mathbf{C}|^2})/2),
\label{eq:entropy}
\end{equation}
where
\[ h(x)=-x \log_{2}x-(1-x)\log_2 (1-x).\]

On the other hand, one obtain $\rho_{B}$ by tracing out the degree
of freedom of part A, i.e.,
\begin{equation}\label{roub}
\rho_{B}=a^{\dagger}a.
\end{equation}
This is a $3\times3$ matrix whose eigenvalues are denoted by $\tilde{\kappa}_{1}^2$,
$\tilde{\kappa}_{2}^2$, $\tilde{\kappa}_{3}^2$ are roots of the algebraic equation
\begin{equation}
\lambda^3-\lambda^2+\frac{|\mathbf{C}|^2}{4}\lambda-\det(\rho_{B})=0.
\end{equation}
The reduced density matrix
$\rho_{B}$ is of rank 2, i.e., $\det\rho_{B}=0$,
then there are only two non-zero eigenvalues.
The von Neumann
entropy takes the same form
as in
Eq.(\ref{eq:entropy}). Just like the case of
Wootters\cite{wootters}, the von Newmann entropy here is aslo a
monotonous function of the norm of concurrence
vectors: $|{\bf C}|^2$.
Therefore, the concurrence vector is a reliable measurement of entanglement
of the states of qubit-qutrit system.

\section{Average concurrence of a Hilbert subspace}\label{sec:average}

As is known that the ground state of physical systems
is degenerate frequently. We need a
definition of the entanglement for Hilbert
subspace to evaluate the entanglement of the degenerate ground state.
With the help of the concurrence vector
applicable to measure the entanglement of individual
state, we suggest to use \emph{average concurrence}.
Since a general state in a Hilbert
subspace can be expanded in terms of its bases, the
magnitude of concurrence vector depends on the
coefficients in the state expansion. The
normalization condition gives a restriction on
the coefficients so that the parameter space of the
Hilbert subspace manifests a compact hyper surface. It
is therefore natural to define average concurrence by the following ratio,
\begin{equation}
\mathcal{C}_{av}=\frac{\int d\mu(p_1, p_2, ...)|\mathbf{C}(p_1, p_2, ...)|}
   {\int d\mu(p_1, p_2, ...)}.
\label{eq:averageE}
\end{equation}
Here $d\mu(p_1, p_2, ...)$
refers to the Haar measure
with respect to the
parametrization $p_1,
p_2,...$, which is
invariant under unitary
operations. For doubly
degenerate case, a general
state can be described by a
superposition of  two
states $\ket{\psi_1}$ and
$\ket{\psi_2}$, the
parameter space is a three
dimensional sphere $S^3$.
The evaluation of average concurrence becomes a calculation of the
integrals in Eq.~(\ref{eq:averageE}). We will
apply the average concurrence to discuss the
ground state of a concrete model in next section.

\section{The ground state superpositions}\label{sec:superposition}

We consider a system of spin 1 and 1/2 with
anisotropic Heisenberg coupling in an uniform magnetic field,
\begin{equation}
 H=\frac{J}{2} (\sigma_x\cdot S_{x}+\sigma_y\cdot S_y
  +\Delta \sigma_z\cdot S_z )+B(\frac{1}{2}\sigma_z+S_z)
  \label{eq:hamiltonian}
\end{equation}
where $\sigma$'s refer to the Pauli matrices for spin-$1/2$
and $S$'s denote the spin operators for spin-$1$;
$J$ stands for their coupling strength, and $\Delta$ represents
the anisotropy of the coupling. Throughout this paper, the
spin-$1/2$ states are denoted by $\ket{\uparrow}$ and
$\ket{\downarrow}$, while the spin-$1$ states are denoted by
$\ket{\Uparrow}$, $\ket{0}$, and $\ket{\Downarrow}$. In terms of the spin-1
matrices
($\hbar$ is put to unit in this paper), the Hamiltonian
(\ref{eq:hamiltonian}) is written out in matrix form:
\begin{equation}
\left(
\begin{array}{cccccc}
  \Delta \frac{ J}{2}+ \frac{3}{2} B & 0 & 0 & 0 & 0 & 0 \\
  0 & \frac{  B}{2} & 0 & \frac{J  }{\sqrt{2}} & 0 & 0 \\
  0 & 0 & -\frac{B}{2}-\frac{\Delta J  }{2} & 0 & \frac{J  }{\sqrt{2}} & 0 \\
  0 & \frac{ J  }{\sqrt{2}} & 0 & \frac{B}{2}-\frac{\Delta J  }{2} & 0 & 0 \\
  0 & 0 & \frac{J  }{\sqrt{2}} & 0 & -\frac{  B}{2} & 0 \\
  0 & 0 & 0 & 0 & 0 & \Delta \frac{ J}{2}- \frac{3}{2} B \\
\end{array}
\right)
  \nonumber
\end{equation}
which solves six eigenvalues and six eigenstates. Among them, the state
$\ket{\uparrow\Uparrow}$ with eigenenergy $\frac{3}{2}B+\frac{1}{2}\Delta J$ and another
one $\ket{\downarrow\Downarrow}$ with eigenenergy $-\frac{3}{2}B +\frac{1}{2}\Delta J$
are obviously non-entangled. The other four states, whose Schmidt numbers are $2$, are
clearly entangled. For simplicity, we put $J$ to unit from now on.

\subsection{In the absence of magnetic field}

When $B=0$, the diagonalization of the Hamiltonian gives rise to
three distinct eigenvalues:
$\frac{1}{2}\Delta$,
$\frac{1}{4}(-\Delta-\sqrt{8+\Delta^{2}})$, and
$\frac{1}{4}(-\Delta+\sqrt{8+\Delta^{2}})$,
of which each energy level is doubly degenerate. The
ground state energy has a critical point $\Delta_c=-1$.

When $\Delta<-1$, the ground state with energy $\frac{1}{2}\Delta$
is doubly degenerate and a general ground state (not restricted to
be eigenstate) is given by a superposition of those two states:
\begin{equation}
\ket{\Psi_{FM}}=a\ket{\downarrow \Downarrow}
  +b\ket{\uparrow \Uparrow},
\end{equation}
where the coefficients fulfil $|a|^2+|b|^2=1$. The
norm of concurrence vector of this state is $2|ab|$.

When $\Delta>-1$, the ground states are doubly
degenerated whose energy reads
$$\frac{1}{4}(-\Delta-\sqrt{8+\Delta^{2}}\,).$$
A general state in this two-dimensional Hilbert subspace is given by
\begin{eqnarray}
\ket{\Psi_{AF}}=c|\psi_1\rangle+d|\psi_2\rangle
\label{eq:psi}
\end{eqnarray}
with $|c|^{2}+|d|^{2}=1$, where
\begin{eqnarray*}
\ket{\psi_1}&=&\frac{1}{F_{+}}\Bigl(\ket{\downarrow 0}
 -\frac{\Delta+\sqrt{\Delta^2+8}}{2\sqrt{2}}\ket{|\uparrow\Downarrow}\Bigr),\\[1mm]
\ket{\psi_2}&=&\frac{1}{F_{-}}\Bigl(\ket{\downarrow\Uparrow}
  +\frac{\Delta-\sqrt{\Delta^2+8}}{2\sqrt{2}}\ket{\uparrow 0}\Bigr),
\end{eqnarray*}
and
$$
F_{\pm}=\sqrt{\bigl(\frac{\Delta\pm\sqrt{8+\Delta^2}}{2\sqrt{2}}\bigr)^2 +1}.
$$
The norm of concurrence vector for this ground state is
\begin{equation}
|\textbf{C}_{\ket{\psi_{AF}}}|=\bigl|\frac{2\bigl(4c^4+4d^4+c^2d^2(4+\Delta^2
 +\Delta\sqrt{8+\Delta^2})\bigr)}{8+\Delta^2}\bigr|^{1/2}.
\label{eq:theC}
\end{equation}

\begin{figure}
\includegraphics[width=6.6cm]{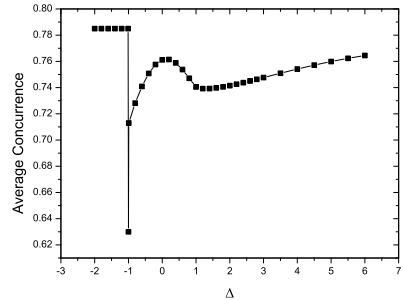}
\caption{\label{fig:b0}
The average concurrence for $B=0$.}
\end{figure}

The point $\Delta=-1$ is a special point in the
absence of magnetic field because the ground state is
4-fold degenerated then. The general state at that point reads
\begin{equation}
\ket{\Psi_c}
   =a\ket{\downarrow\Downarrow}+b\ket{\uparrow\Uparrow}
   +c\ket{\phi_1} + d\ket{\phi_2},
\end{equation}
where
\begin{eqnarray}
\ket{\phi_1} &
=&\sqrt{\frac{2}{3}}\ket{\downarrow0}
-\sqrt{\frac{1}{3}}\ket{\uparrow\Downarrow}
       \nonumber\\
\ket{\phi_2} &
=&\sqrt{\frac{1}{3}}\ket{\downarrow\Uparrow}
-\sqrt{\frac{2}{3}}\ket{\uparrow
0}.
\label{eq:phi}\end{eqnarray}
The norm of its concurrence vector is given by
\begin{eqnarray*}
\lefteqn{|\mathbf{C}(a,b,c,d)|=\bigl|4a^{2}b^{2}+\frac{8}{3}(b^{2}c^{2}+a^{2}d^{2}+a
b c d)}\\&
&+\frac{8}{9}(c^{4}+d^{4})
 -\frac{16}{3\sqrt{3}}cd(ac-bd)+ \frac{4}{9}c^{2}d^{2}\bigr|^{1/2}.
\end{eqnarray*}
and the normalization condition requires $|a|^{2}+|b|^{2}+|c|^{2}+|d|^{2}=1$. This
reveals that the parameter space is a 7-dimensional sphere $S^7$. The average
concurrence can be calculated by means of Monte Carlo method. As the result, the average
concurrence versus $\Delta$ is plotted in Fig.~\ref{fig:b0}. Obviously, the average
concurrence is discontinuous (suddenly drops to 0.62) at the phase transition point
$\Delta=-1$ and it reaches a local maximum value $0.76$ at $\Delta=0.1$, and approaches
to $0.785$ when $\Delta\rightarrow\infty$.

\subsection{In the presence of magnetic field}

It is known that the application of an external field will break
down the symmetry of the model, which brings about splitting of
degenerate energy levels. Consequently, the entanglement of the
system will change. The application of an external magnetic field
along $z$ direction makes the concurrence varies as a function of
the anisotropic parameter $\Delta$ and external field B.

When $\Delta\leq-1$, the ground state is $\ket{\uparrow\Uparrow}$ with energy
$\Delta/2+3B/2$ for diamagnetic response $B<0$, For paramagnetic response $B>0$, it is
$\ket{\downarrow\Downarrow}$ whose energy is $\Delta/2-3B/2$. Obviously the state in
both cases are not entangled. However, for $B=0$, the ground state is doubly degenerate,
\begin{equation}
c_{1}\ket{\uparrow\Uparrow}+c_{2}\ket{\downarrow\Downarrow},
\end{equation}
of which the norm of concurrence vector is
$2|c_{1}c_{2}|$. Then the average concurrence is
calculated to be $0.785$. Meanwhile we find that the concurrence has one singular
point along the $B$-parameter in the ferromagnetic regime $\Delta<-1$.

\begin{figure}
\includegraphics[width=4.6cm]{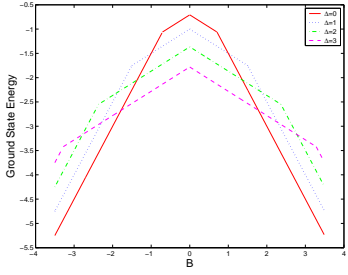}
\caption{\label{fig:e-b}
Ground state energy versus B that indicates the critical point
where level crossing occurs.}
\end{figure}

\begin{figure}
\setlength{\unitlength}{0.01mm}
\includegraphics[10cm,0cm][0cm,8cm]{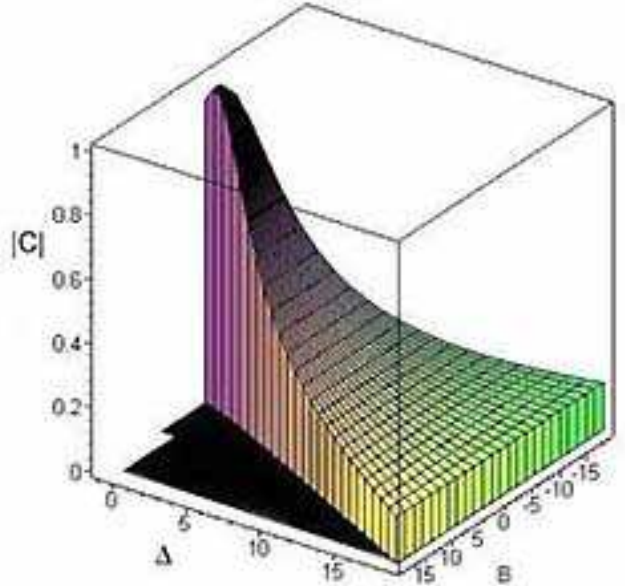}
\caption{\label{fig:c-b}
The norm of concurrence vector versus the magnetic field and
anisotropic parameter.}
\end{figure}

When  $\Delta>-1$, the plot of energy versus $B$ (fig.\ref{fig:e-b}) exhibits that there
are three critical points. When $B<-(3\Delta+\sqrt{8+\Delta^2})/4$, the ground state
$\ket{\uparrow\Uparrow}$ with energy $3B/2+\Delta/2 $ is obviously not entangled. In the
region $-(3\Delta+\sqrt{8+\Delta^2})/4<B<0$, the ground state whose energy takes
$(2B-\Delta-\sqrt{8+\Delta^{2}})/4$ becomes
\begin{equation}\label{aa}
 \frac{1}{N_{1}}(\frac{\Delta-\sqrt{8+\Delta^{2}}}{2\sqrt{2}}\ket{\uparrow0}+\ket{\downarrow\Uparrow}),
\end{equation}
where $N_{1}$ denotes the normalization factor
\[
N_{1}=\Bigl(\bigl(\frac{\Delta-\sqrt{8+\Delta^{2}}}{2\sqrt{2}}
 \bigr)^{2}+1\Bigr)^{\frac{1}{2}}.
\]
The norm of concurrence vector of state (\ref{aa}) is obtained
\begin{equation}
C(\rho)=\frac{4\sqrt{2}(\sqrt{8+\Delta^2}-\Delta)}{8+(\sqrt{8+\Delta^2}-\Delta)^2}.
\label{cc}
\end{equation}
When $\Delta=0$, the entanglement reaches
maximum 1. Furthermore, when $0<B<(3\Delta+\sqrt{8+\Delta^2})/4$, the ground state with
energy $(-2B-\Delta-\sqrt{8+\Delta^{2}})/4$ is given by
\begin{equation}
\frac{1}{N_{2}}\Bigl(-\frac{\Delta+\sqrt{8+\Delta^{2}}}{2\sqrt{2}}\ket{\uparrow\Downarrow}
+\ket{\downarrow0},
\Bigr)
\end{equation}
where $N_{2}$ is normalization factor. The norm of concurrence vector is the same as
Eq.~(\ref{cc}). Once $B>(3\Delta+\sqrt{8+\Delta^2})/4$, the ground state with
energy $-3B/2+\Delta/2$ becomes $\ket{\downarrow\Downarrow}$ which is no more entangled.
As a result, there are three critical points $B=0$, $-(3\Delta+\sqrt{8+\Delta^2})/4$ and
$+(3\Delta+\sqrt{8+\Delta^2})/4$ in the regime $\Delta >-1$.
The norm of concurrence vector as a function of the magnetic field $B$ and anisotropy parameter
$\Delta$ is plotted in Fig.\ref{fig:c-b}. Clearly, when $\Delta$ approaches $-1$,
the width of the peak of the
entanglement curve approaches zero.

\section{The ground state mixtures}\label{sec:mixture}
\begin{figure}
\includegraphics[width=7cm]{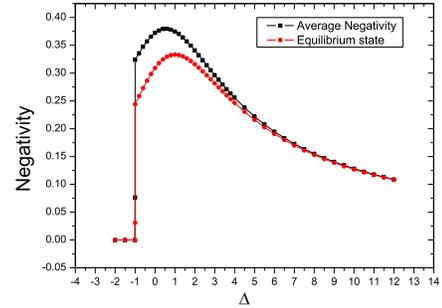}
\caption{\label{mixfinal} Negativity versus $\Delta$ for the equilibrium state at zero
temperature (red) and for the average of general mixture (black) of ground states }
\end{figure}

It is interesting to discuss the entanglement
feature of a mixtures of the degenerate ground
states of the same model. The \emph{negativity}
introduced by G. Vidal~\emph{et al} is known to be
a useful measurement for the entanglement of mixed
states \cite{vidal}, namely,
\begin{equation}
\mathcal{N}(\rho)\equiv\frac{\|\rho^{T_{A}}\|_{1}-1}{2},
\end{equation}
where the trace norm is defined by
$\|A\|_1 \equiv
tr\sqrt{A^{\dagger}A}$
and $T_{A}$ refers to the partial transposition of A.
The negativity vanishes for
unentangled states.

Hereafter we discuss the mixture of ground state in the absence of
magnetic field. In the ferromagnetic regime, $\Delta <-1$, a general
mixed state is given by
\begin{equation}
\rho=p|\downarrow\Downarrow\rangle\langle\downarrow\Downarrow|
   +(1-p)|\uparrow\Uparrow\rangle\langle\uparrow\Uparrow|,
\label{state1}
\end{equation}
Because the convexity of the norm of concurrence vector, one can obtain
\begin{equation}
C(\rho)\leq\sum_{i}p_{i}C(\rho_{i}).
\end{equation}
Clearly, both the concurrence and negativity of the state described by (\ref{state1}) are zero.

In the antiferromagnetic regime $\Delta > -1$. The density matrix for a general mixture of ground
state is given by \begin{equation}\label{state2}
\rho=p\ket{\psi_1}\bra{\psi_1}
+(1-p)\ket{\psi_2}\bra{\psi_2},
\end{equation}
where $\ket{\psi_1}$ and $\ket{\psi_2}$ were given in Eq.(\ref{eq:psi}), its negativity
is obtained after some algebra,
\begin{equation}
\mathcal{N}(\rho)=\frac{\Delta}{4\sqrt{8+\Delta^2}}-\frac{1}{4}+f(p)+f(1-p)
\end{equation}
where
$$f(p)=\sqrt{\frac{16-32p+(20+\Delta^2-\Delta\sqrt{8+\Delta^2})p^2}{8(8+\Delta^2)}}$$

At the critical point $\Delta=-1$, the density matrix for mixture of ground states becomes
\begin{eqnarray}
\rho&=&p_1|\downarrow\Downarrow\rangle\langle\downarrow\Downarrow|
 +p_2\ket{\uparrow\Uparrow}\bra{\uparrow\Uparrow}+p_3\ket{\phi_1}\bra{\phi_1}
 \nonumber\\
 & &
 +(1-p_1-p_2-p_3)\ket{\phi_2}\bra{\phi_2}.
\label{state3}
\end{eqnarray}
where $\ket{\phi_1}$ and $\ket{\phi_2}$ were given in Eq.(\ref{eq:phi}). The
negativity for the mixture of equilibrium at zero temperature ($p_j=1/4$) is
$0.031$, and its average value is $0.077$.

The negativity versus the anisotropy parameter
$\Delta$ is plotted in Fig.\ref{mixfinal}.
One can see from the plot that there is a singularity
at $\Delta=-1$ where the quantum phase transition
(ferromagnetic to antiferromagnetic) occurs.
Whereas, when $\Delta$ increases from $-1$, the negativity  rises at first then descends
after reaching a maximal value. Finally it approaches to zero when $\Delta$ goes to infinity.
The state with $p=1/2$ is particularly interesting because it can
be regarded as the thermal equilibrium at zero temperature. Note that the
negativity of equilibrium state at zero temperature reaches $1/3$ at $\Delta=1$
where the system recovers its largest symmetry (isotropic Heisenberg coupling).
Similar features have been noticed in some other models~\cite{GuLinLi,GuLiLin}

\section{Summary and discussion}

We have studied the entanglement feature of the ground state for system of spin $1$ and
$1/2$. We have shown that the concurrence vector is consistent with the measurement of
von Neumann entropy for such system. Because its ground state is degenerate in cases,
the simple calculation of norm of concurrence for a state is no more applicable. We
therefore proposed a concept, \emph{average concurrence}, to measure the entanglement of
Hilbert subspace. Based on this definition, we discussed the entanglement of the
superposition of the degenerated ground states. We obtained the relations between the
average concurrence and the anisotropy parameter. We also studied the model by taking
account of external magnetic field. The relation between the norm of concurrence and the
magnetic field and the anisotropy parameters are calculated.  We found
that the state is not entangled when anisotropy factor $\Delta<-1$. When $\Delta
>-1$, the concurrence varies with respect to the anisotropy factor
$\Delta$. We also studied the entanglement of a general mixture of the degenerate ground
state by employing the widely used negativity.

Our results indicate that the averages of both concurrence and negativity
have singularities at the quantum critical point $\Delta=-1$. The negativity
for the equilibrium at zero temperature reaches the maximal value at
$\Delta=1$ where the model possesses the largest symmetry. However, both the negativity
averaged over the general mixture and the norm of concurrence vector averaged over
the general superposition of the degenerate ground states do not reach maximum at $\Delta=1$.
The average concurrence takes the largest value $0.785$ in the ferromagnetic
regime $\Delta<-1$ and in the limit of antiferromagnetic Ising dominant regime
$\Delta\rightarrow\infty$.

This work is supported by NSFC No.10225419 and 90103022. Helpful
discussions with X.G. Wang are acknowledged.

\end{document}